\documentclass[twocolumn]{aastex631}

\usepackage{nccmath}
\usepackage{xcolor}
\usepackage{hyperref}
\usepackage[anythingbreaks]{breakurl}
\usepackage{graphicx}
\newcommand{\degree}{$^\circ$}

\usepackage{soul}
\usepackage{tabularx}
\usepackage[shortlabels]{enumitem}
                    \setlist[enumerate, 1]{1\textsuperscript{o}}

\begin{document}

\title{Power-law Distribution of Solar-Cycle Modulated Coronal Jets}

\correspondingauthor{Jiajia Liu, Yuming Wang, Robertus Erd{\'e}lyi}
\email{ljj128@ustc.edu.cn, ymwang@ustc.edu.cn, robertus@sheffield.ac.uk}

\author[0000-0003-2569-1840]{Jiajia Liu}
\affiliation{Deep Space Exploration Lab/School of Earth and Space Sciences, University of Science and Technology of China, Hefei, 230026, China}
\affiliation{Astrophysics Research Centre, School of Mathematics and Physics, Queen's University, Belfast, BT7 1NN, UK}
\affiliation{CAS Key Laboratory of Geospace Environment, Department of Geophysics and Planetary Sciences, University of Science and Technology of China, Hefei, 230026, China}

\author[0000-0002-3237-3819]{Anchuan Song}
\affiliation{CAS Key Laboratory of Geospace Environment, Department of Geophysics and Planetary Sciences, University of Science and Technology of China, Hefei, 230026, China}
\affiliation{CAS Center for Excellence in Comparative Planetology, University of Science and Technology of China, Hefei, 230026, China}

\author[0000-0002-9155-8039]{David B. Jess}
\affiliation{Astrophysics Research Centre, School of Mathematics and Physics, Queen's University, Belfast, BT7 1NN, UK}
\affiliation{Department of Physics and Astronomy, California State University Northridge, Northridge, CA 91330, USA}

\author[0000-0003-0951-2486]{Jie Zhang}
\affiliation{Department of Physics and Astronomy, George Mason University, Fairfax, VA 22030, USA}

\author[0000-0002-7725-6296]{Mihalis Mathioudakis}
\affiliation{Astrophysics Research Centre, School of Mathematics and Physics, Queen's University, Belfast, BT7 1NN, UK}

\author[0000-0002-3606-161X]{Szabolcs So{\'o}s}
\affiliation{Department of Astronomy, E\"{o}tv\"{o}s Lor\'{a}nd University, Budapest, P\'{a}zm\'{a}ny P. s\'{e}t\'{a}ny 1/A, H-1117, Hungary}

\author[0000-0001-5435-1170]{Francis P. Keenan}
\affiliation{Astrophysics Research Centre, School of Mathematics and Physics, Queen's University, Belfast, BT7 1NN, UK}

\author[0000-0002-8887-3919]{Yuming Wang}
\affiliation{Deep Space Exploration Lab/School of Earth and Space Sciences, University of Science and Technology of China, Hefei, 230026, China}
\affiliation{CAS Key Laboratory of Geospace Environment, Department of Geophysics and Planetary Sciences, University of Science and Technology of China, Hefei, 230026, China}
\affiliation{CAS Center for Excellence in Comparative Planetology, University of Science and Technology of China, Hefei, 230026, China}

\author[0000-0003-3439-4127]{Robertus Erd{\'e}lyi}
\affiliation{Solar Physics and Space Plasma Research Centre (SP2RC), School of Mathematics and Statistics, The University of Sheffield, Sheffield, S3 7RH, UK}
\affiliation{Department of Astronomy, E\"{o}tv\"{o}s Lor\'{a}nd University, Budapest, P\'{a}zm\'{a}ny P. s\'{e}t\'{a}ny 1/A, H-1117, Hungary}
\affiliation{Gyula Bay Zolt\'an Solar Observatory (GSO), Hungarian Solar Physics Foundation (HSPF), Pet\H{o}fi t\'er 3., Gyula, H-5700, Hungary}

\begin{abstract}

Power-law distributions have been studied as a significant characteristic of non-linear dissipative systems.  Since discovering the power-law distribution of solar flares that was later extended to nano-flares and stellar flares,  it has been widely accepted that different scales of flares share the same physical process. Here, we present the newly developed Semi-Automated Jet Identification Algorithm (SAJIA) and its application for detecting more than 1200 off-limb solar jets during Solar Cycle 24. Power-law distributions have been revealed between the intensity/energy and frequency of these events, with indices found to be analogous to those for flares and coronal mass ejections (CMEs). These jets are also found to be spatially and temporally modulated by the solar cycle forming a butterfly diagram in their latitudinal-temporal evolution, experiencing quasi-annual oscillations in their analysed properties, and very likely gathering in certain active longitudinal belts.  Our results show that coronal jets display the same nonlinear behaviour as that observed in flares and CMEs, in solar and stellar atmospheres, strongly suggesting that they result from the same nonlinear statistics of scale-free processes as their counterparts in different scales of eruptive events. Although these jets, like flares and other large-scale dynamic phenomena, are found to be significantly modulated by the solar cycle, their corresponding power-law indices still remain similar.

\end{abstract}

\section{Introduction} \label{sec:intro}

Solar flares were found to exhibit a power-law distribution of their frequency \citep[e.g.,][]{Dennis1985,Crosby1992}. This power-law distribution was later found to also apply to nano-flares \citep{Benz1999} and stellar flares \citep{Maechara2015,Shibayama2013}. In addition to observations, continuous efforts have also been made in modelling \citep[e.g.,][]{Lu1991, Hamon2002} to explain these power-law distributions. Though the physics behind power-law distributions needs more exploration, it is widely accepted that they are essential evidence that different scales of flares are triggered by the same underlying physical process, i.e.,  large flares are avalanches of a number of (smaller-scale reconnection) events \citep{Lu1991}, similar to avalanches of sand piles. A natural question would then be: do other types of localised eruptive events share the same physical mechanism as flares in the solar atmosphere?

Though less intense than flares, one of the common localised phenomena in the solar atmosphere is coronal jets. Coronal jets are multi-thermal elongated eruptive events detected in the corona in various wavelengths including UV/EUV and x-rays \citep[see reviews in e.g.,][]{Shibata1992,Raouafi2016}. Most theories and observations suggest that coronal jets are triggered by magnetic reconnection between closed and open magnetic field lines \citep[e.g.,][]{Shibata1992,Canfield1996,Moore2010,Pariat2015,Sterling2015}. Besides the vastly different plasma properties at the locations where the magnetic reconnection happens \citep{Shibata2007},  the process usually occurs in regions where magnetic field lines are closed for intensive flares \citep[e.g.,][]{Lin2000}, whereas totally different magnetic topologies are exhibited for coronal jets. The study of coronal jets is of great interest mainly due to the associated wide variety of physical processes (e.g., magnetic reconnection, MHD instabilities and waves) involved during their eruptions \citep{Raouafi2016} and the important role of their chromospheric analogs (i.e., spicules) in atmospheric heating and solar wind acceleration \citep[e.g.,][]{Pontieu2004,Shibata2007,Tian2014}.

On the other hand, the Sun, like many other stars, undergoes a number of cycles that have a significant impact on the solar system including the habitability of its planets \citep{Brehm2021}. The best-known and studied cycle is the 11-year activity cycle, evidenced by a number of global phenomena including the long-term evolution of sunspot numbers, irradiance variations, solar flares, and CMEs \citep{Solanki2011,Song2016,Bhowmik2018}. One of the most prominent features is the so-called ``butterfly diagram'' of sunspots and active regions \citep{Maunder1904}. Other variations include the quasi-annual/biennial oscillation and longer ($\sim$hundred years) cycles \citep{Hathaway1999,McIntosh2015,Gyenge2016}. Though solar cycles have been widely suggested as a marker of the solar dynamo \citep{Parker1955,Charbonneau2010} i.e. the alternating conversion between the global large-scale poloidal and toroidal fluxes, it is still unclear how and to what extent the solar cycle could modulate the behaviour of localised small-scale solar features, e.g., coronal jets.

To proceed in answering the above questions, statistical studies of a large number of jets are needed. However, identifying localised jets among the wealth of solar features is not a trivial task owing to their relatively faint, short-lived, and small appearance. Thus many relevant works have been done on manually identifying jets. For example, \cite{Liu2016b} investigated the thermal, kinetic, and magnetic properties of 11 homogeneous coronal jets and found that stronger magnetic reconnections tend to trigger larger coronal jets with more thermal and kinetic energies.  Manually checking the X-ray observations provided by the Yohkoh Soft X-Ray Telescope \citep{Ogawara1995}, \cite{Shimojo1996} identified 100 jets that mostly originated from active regions during the period from November 1991 to April 1992. Statistical properties of these jets were explored, and a power-law distribution was found between the peak intensity of jets' footpoints and their frequency of appearance. However, the flatter power-law index (-1.2) they found was in doubt due to the limited number of events, compared to the power-law indices found from nanoflares and microflares \cite[e.g., -1.74 and -1.79 by][respectively]{Shimizu1995,Aschwanden2000}.

In this paper, we present a novel semi-automated identification algorithm of off-limb coronal jets - SAJIA, and its application to observations by the Atmospheric Imaging Assembly \citep[AIA,][]{Lemen2012} onboard the Solar Dynamics Observatory (SDO) during Solar Cycle 24 from 2010 to 2020. Power-law distributions are found between the intensity/energy and frequency of the 1215 jets that were identified using SAJIA. A butterfly diagram indicates the migration of jets' formation loci from higher latitudes to lower ones, and quasi-annual oscillations of their studied properties are revealed. The structure of this paper is organised as follows: the dataset and the Semi-Automated Jet Identification Algorithm (SAJIA) are presented in Sec.~\ref{sec:dam}. Results are exhibited in Sec.~\ref{sec:result}, followed by discussions and conclusions in Sec.~\ref{sec:conc}.

\section{Data and Method} \label{sec:dam}
\subsection{Data}
The Atmospheric Imaging Assembly \citep{Lemen2012} (AIA) on-board the Solar Dynamics Observatory (SDO) provides simultaneous full-disk images of the Sun at eight narrow EUV passbands covering the temperature range from $6\times10^4$ K to $2\times10^7$ K with a pixel size of $\sim$0.6\arcsec\ and a cadence of 12 s. Full-disk images observed at the 304 \AA\ passband from 2010 June 1 to 2020 May 31 at 00 UT, 06 UT, 12 UT, and 18 UT on each day are used for this work. Each image has an original size of 4096$\times$4096 pix$^2$. The 304 \AA\ passband was chosen because it has a much less bright off-limb background, compared to other AIA passbands, making it easier for automated identification algorithms to extract jets from the background.

\subsection{Semi-Automated Jet Identification Algorithm (SAJIA)}
 For a given SDO/AIA 304 \AA\ full-disk observation, the solar disk is masked from its centre to 0.02 solar radius (i.e. $\sim$14 Mm) above the limb to avoid the forest of spicules \citep{Pereira2012}, because we focus on the detection of off-limb coronal jets. A corresponding background image is then constructed using the mean of four images at 02 UT, 08 UT, 14 UT, and 20 UT on the same day, and the mean values are then subtracted from the target image. The image size is then downgraded to 512$\times$512 pix$^2$ to reduce the computational power required. This downgraded image is further normalised to [0, 255] by setting the lower limit as the average value ($av$) of all valid off-limb pixels. The image is then further converted to a binary image by using a lower limit ($ll$) of 120. In addition to the consideration that coronal jets are usually brighter than the background, $av$ and $ll$ were chosen via trial and error by testing a number of random images encompassing the full extent of the time series. This made it possible to validate that SAJIA could consistently isolate the features of interest irrespective of the variation of the overall brightness of the Sun over the solar cycle.

Applying the Douglas-Peucker algorithm \citep{Douglas1973}, which is one of the most widely used algorithms to decimate a curve to a similar one with fewer nodes, to all contours found from the binary image, a series of polygons with the least edges representing their corresponding closed contours are obtained. Considering that most off-limb jets are simple elongated structures resembling rectangles in the binary image, only polygons with four edges rooted at the solar limb are kept. To avoid large-scale filaments usually seen horizontally off the limb, polygons with inclination angles (with respect to the local radial directions) larger than 60\degree\ are further removed. We then only keep polygons with aspect ratios greater than 1.5, recalling that a coronal jet should have an aspect ratio larger than three \citep{Shimojo1996}.

The remaining polygons after the above processes are kept as candidates for off-limb coronal jets. Parameters of these candidates are then calculated with their intensities corrected after considering the instrument degradation \citep{Barnes2020}. We note that apart from the aspect ratio determined by previous observations, there were two important independent arbitrary parameters that may affect the detection performance: the lower limit used for the normalisation and the upper limit of the inclination angle. For this purpose, the lower limit of the normalisation angle was applied to the entire image instead of a particular part, meaning it should only affect the number of events detected but not treat jets at different locations differently. In regard to the upper limit of the inclination angle, it should only affect the number of events detected but not the distribution of jet locations, considering that off-limb open magnetic field lines are generally not far away from the local radial direction.

\begin{figure}[htb]
\centering
\includegraphics[width=\hsize]{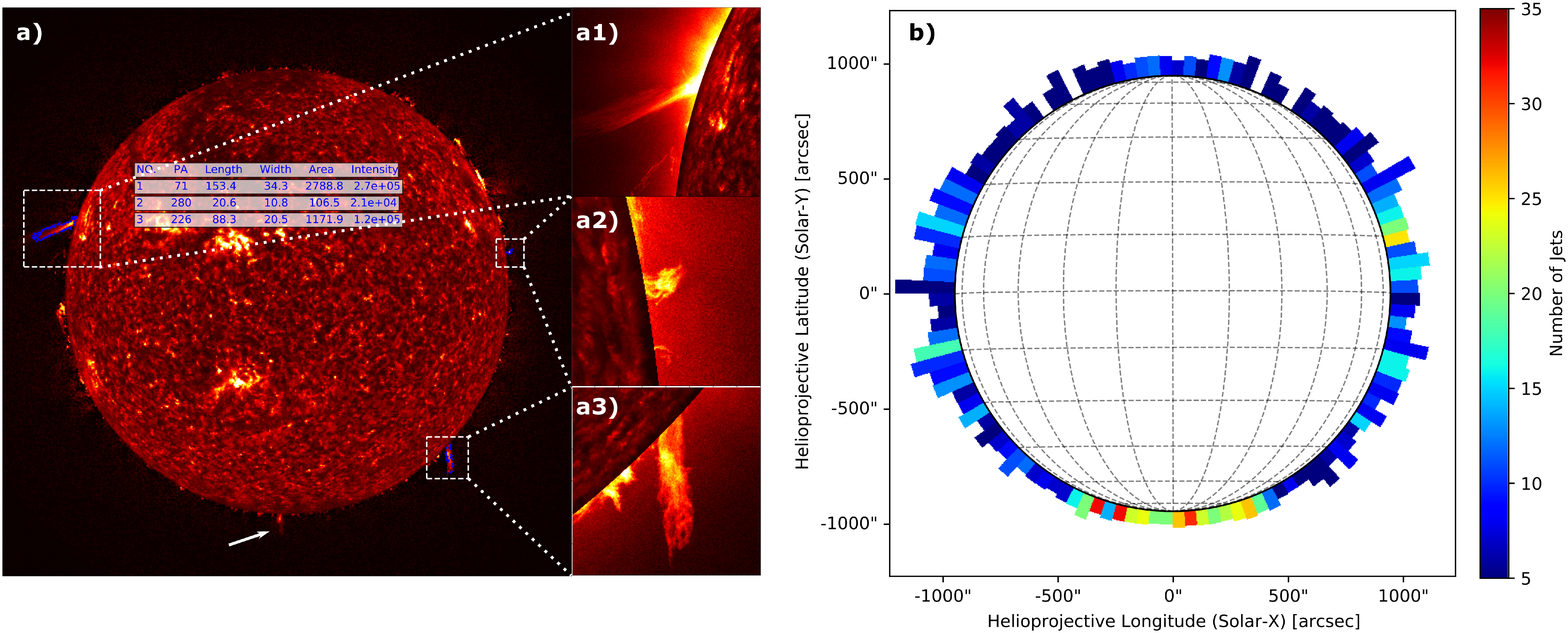}
\caption{\textbf{An example and statistics of the detected off-limb coronal jets.} \textbf{a)} The Sun observed in the SDO/AIA 304 \AA\ passband at 06:05 UT on 2010 June 27. Detected jet candidates are enclosed in blue curves. The white arrow points to an irregular jet that was missed by the automated detection algorithm. Texts in the image provide the calculated properties of the three candidates with their position angles (PA) in degrees, plane-of-the-sky projected length in Mm, width in Mm, area in Mm$^2$ and total intensity in digital numbers from the original observation before correcting the instrument degradation. \textbf{a1) - a3)} are the zoom-in views of the detected jet candidates. An animation of panels a1) - a3) is available to show the temporal evolution. These images were taken by SDO/AIA at its 304 \AA\ passband on 2010 June 27. The animation begins at 05:30:03 and ends at 06:29:03. The animation's real-time duration is 3 seconds. The white circle in \textbf{b)} represents the solar disk with dotted lines drawn in 15\degree\ intervals. The length and colour of each coloured bar are twice the average plane-of-the-sky projected length and number of jets detected in every 3\degree\ interval, respectively.}
\label{fig_example}
\end{figure}

After obtaining information on all candidates from the above procedure, each of them was validated manually and non-jet events were removed. During the manual validation process, jets were identified as collimated plasma ejections shooting upward into the corona with enhanced emission. The advantage of the above laborious semi-automated process is that it treats all images and events equally with a single self-consistent selection procedure to minimise human bias. Figure~\ref{fig_example}a) is an example of the off-limb jet candidates detected by SAJIA in the 304~\AA\ passband of the SDO/AIA at 06:05 UT on 2010 June 27. Three candidates are enclosed within blue curves outlining their edges with their calculated properties given in the inset. The jet on the left (Fig.~\ref{fig_example}a1) has been successfully detected while a prominent one marked by the white arrow was missed. The missed detection is due to its irregular shape. The two candidates on the right are most likely a straight prominence (which did not erupt, Fig.~\ref{fig_example}a3) and a mini-filament eruption (which appeared as a curved arcade instead of a collimated eruption, Fig.~\ref{fig_example}a2), respectively. Thus, in spite of their appearance, they are not real jets (see the animation of panels a1 - a3 for their temporal evolution). 

SAJIA was applied to the SDO/AIA 304 \AA\ full-disk observations at four different times (00 UT, 06 UT, 12 UT and 18 UT) daily from 2010 June 01 to 2020 May 31. We then manually investigated every candidate and removed false detections and confirmed 1215 coronal jets. As illustrated above, parameters of jets including their position angle (PA), length, width, area, and intensity are given automatically by SAJIA, and available at \href{http://space.ustc.edu.cn/dreams/sajia}{http://space.ustc.edu.cn/dreams/sajia}. To estimate the missing rate of the above process, 2 months were randomly chosen (August 2010 and April 2014), and off-limb jets in the resulting 245 images were manually identified. It was found that at least 22 coronal jets in low latitudes (between $\pm$60\degree) were missed (August 2010: 9 jets; April 2014: 13 jets), with 52 detected by SAJIA. The missed jets, identified manually, were not added to the list to keep the consistency of the whole dataset. This gives an approximate overall missing rate of 30\% in non-polar regions. It was difficult to estimate the missing rate in polar regions due to the existence of a large number of faint events. Owing to the faint nature of coronal jets in polar regions (see Sec.~\ref{sec:result}) and considering the automated detection of jets is heavily affected by their brightness, we suggest that a large number of faint events could have been missed in polar regions.

It is worth noting that the above dataset was built from the detection of coronal jets at a single wavelength (SDO/AIA 304 \AA). This raises the question of whether the events are typical of general coronal jets. However, it is widely argued in the literature that, coronal jets are multi-thermal \citep{Liu2016b,Joshi2020}. One should bear in mind that coronal jets could simultaneously appear at other wavelengths when observed at one particular passband. To verify the representativeness of events in our dataset, we have randomly chosen 20 coronal jets (see Table~\ref{tb_visi}) with their total intensities from the minimum ($\sim$6$\times 10^3$ DN) to the maximum ($\sim$4$\times 10^6$ DN). It is shown that events with total intensities less than 10$^4$ DN (163 events in the whole dataset) are hardly visible in the AIA 171 \AA\ (0.6 MK) and 193 \AA\ (1.6 MK) passbands \citep{Lemen2012}. Events with total intensities between 10$^4$ and $10^5$ DN are mostly seen as bright features in the AIA 171 \AA\ passband and dark features in the AIA 193 \AA\ passband. Events with total intensities between 10$^5$ and 6$\times 10^5$ DN are mostly observed as bright features in both the AIA 171 \AA\ and AIA 193 \AA\ passbands, while events with total intensities above that could be observed in most of the AIA EUV passbands. The above results show that the 304 \AA\ passband is the wavelength for which coronal jets with a wide range of temperatures are best identified.

\section{Results} \label{sec:result}
\subsection{Power-law Distribution} \label{sec:powerlaw}
\begin{figure}[htb]
\centering
\includegraphics[width=\hsize]{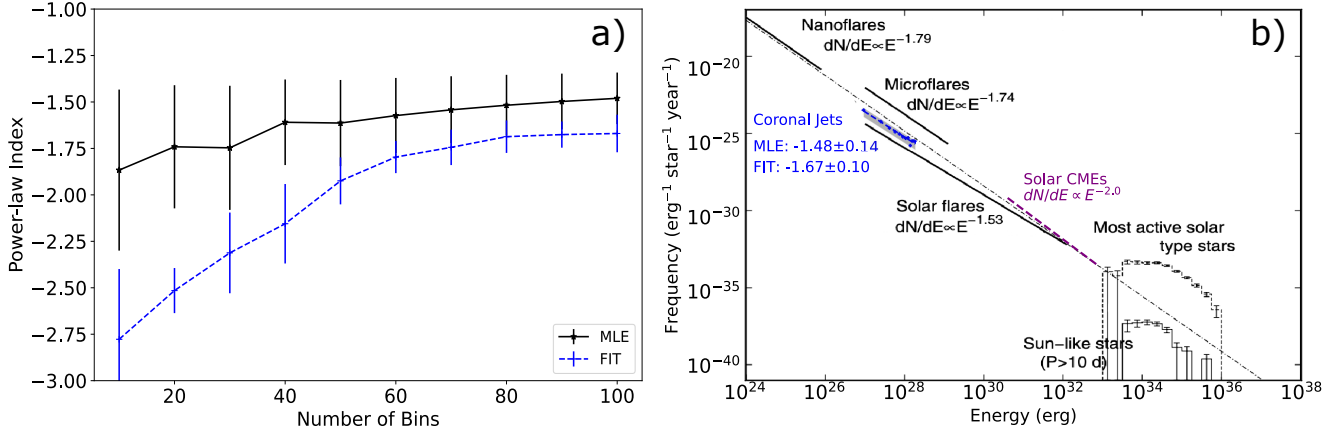}
\caption{\textbf{Power-law index of solar coronal jets.} \textbf{a)} depicts the evolution of the power-law index of the detected off-limb jets obtained from the maximum likelihood estimation (MLE) and least-square fit (dashed curve) with the number of bins varying from 10 to 100. Vertical lines are the corresponding errors. Blue plus signs in \textbf{b)} show the histogram of the log-scale energy-frequency distribution of the detected off-limb jets. The power-law indices estimated from the MLE and least-square fit (FIT) are also shown. The dashed blue line marks the result from the least-square fit, with the grey shadow as the estimated uncertainties of the frequency propagated from the uncertainties (60\%) of the thermal energies as detailed in the main text. Black curves and lines reveal the energy-frequency distributions of solar and stellar flares over a broad energy range, adapted from Fig. 9 in \cite{Shibayama2013}. The black dash-dotted line with a power-law index of -1.8 is the average power-law distribution of all flares from the Sun to Sun-like stars. The purple dashed line shows the energy-frequency distribution of solar coronal mass ejections (CMEs), adapted from Fig. 16i) in \cite{Aschwanden2016}.}
\label{fig_power}
\end{figure}

A collective view of the locations and plane-of-the-sky lengths of all the 1215 off-limb solar coronal jets is shown in Figure~\ref{fig_example}b).  Given the nature of the optically thick emission of the SDO/AIA 304 \AA\ observations, and assuming that the observed jets have axial symmetry, the 304 \AA\ total intensity of an off-limb coronal jet would to some extent reflect its mass and thus the thermal energy \citep{Liu2016b}. A power-law relationship between the total intensity and number density (i.e., frequency) of the detected jets should exist if different scales of jets result from the same nonlinear statistics of scale-free processes which has no characteristic spatial scale and is triggered by a critical instability threshold, e.g., the kink instability \citep{Pariat2015}. Figure~\ref{fig_power}a) shows the variation of the power-law index estimated with the least-square fit (dashed line) and  maximum-likelihood estimation \citep[MLE, solid line in panel a, see][for details of the MLE method]{Cam1990} applied to all 1215 jets, by varying the number of bins from $10$ to $100$. The power-law indices obtained from the least-square fit drop quickly from $-2.3$ and seem to converge between $-1.6$ and $-1.7$. The power-law indices obtained from the MLE do not experience much change and are quite stable between $-1.4$ and $-1.6$. The difference in sensitivity to the bin size between the least-square fit and the MLE approach may be due to the input intensity distribution not strictly following Gaussian statistics, which the MLE approach is able to robustly account for \citep{Huys2016, Jess2019}. These power-law distributions are a strong indication of the self-organised criticality of the triggering magnetic field and coincide well with their counterparts found for solar flares both in modelling \citep{Lu1991} and measurements \citep{Crosby1992}, as well as for observations of super flares on solar-type stars \citep{Maechara2015}.

To directly compare the above power-law distribution of jets with those of flares and CMEs \citep[e.g.,][]{Aschwanden2016}, the energy of jets needs to be estimated. Given the optically thick nature of the SDO/AIA 304 \AA\ passband, we assume that the thermal energy ($E$) of a jet is proportional to its total intensity ($I$), i.e., $E \propto I$. The average thermal energy of solar coronal jets is about $10^{28}$ erg \citep{Shimojo2000}. The thermal energy of each detected coronal jet can then be estimated as $E=10^{28} I/I_{1/2}$ erg. $I_{1/2}$ is defined by 

\begin{equation}
I_{1/2} = 2^{1/(\alpha-1)}I_m,
\label{eq_i}
\end{equation}

\noindent where $I_m$ is the median value of the total intensity of all detected jets, $\alpha$ is the absolute value of the power-law index (5/3), and $2^{1/(\alpha-1)}$ is the median factor for a power-law distribution from \cite{Newman2005}. The following equation is used to make an order-of-magnitude estimate of the frequency of jets:
\begin{equation}
df =  \frac{dN}{dE}\cdot\frac{2\pi}{4\theta} \cdot \frac{t}{\overline{T}} \cdot \frac{1}{1-a} \ ,
\label{eq_df}
\end{equation}

\noindent where $df$ is the frequency of occurrence of all coronal jets in units of erg$^{-1}$ star$^{-1}$ year$^{-1}$, $dN$ is the number of off-limb jets detected each year in each energy bin $dE$, $t$ is the image cadence used for the jet detection algorithm (6 hours), $\overline{T}$ is the average lifetime of coronal jets \citep[$\sim$10 minutes,][]{Shimojo2000}, $a$ is the estimated overall missing rate (0.3) of the detection, and $\theta$ is the maximum angle in longitude away from the solar limb from where a jet could reach above the mask applied (1.02 solar radii), given by:
\begin{equation}
\cos{\theta} = \frac{1.02R_s}{\overline{L} + R_s},
\label{eq_cost}
\end{equation}

\noindent where $\overline{L}$ is the average length ($\sim$38 Mm) of all detected off-limb jets, and $R_s$ is the solar radius. 

In Figure~\ref{fig_power}b), blue plus signs (with a number of events above 5 per bin) show the log-scale distribution of the estimated thermal energy of jets versus the frequency of occurrence (with 100 bins).  The grey shadow shows the estimated uncertainties of the frequency propagated from the
uncertainties of the thermal energy. This is needed to validate the accuracy of the thermal energy power-law distribution and is described in the next two paragraphs. A least-square fit (blue line) reveals a power-law index of $-1.63\pm0.10$, with MLE yielding a less steep power-law index of $-1.42\pm0.14$. In the above, a linear relationship between the 304 \AA\ intensities and the thermal energies of the observed coronal jets was assumed. However, the formation of the 304 \AA\ passband is complex and, as an optically thick one, it experiences additional physical processes (especially photoionisation) than the optically thin passbands.

The highest total intensity of coronal jets used to study the power-law distribution shown as the blue line in Fig.~\ref{fig_power}b) is around 6.0$\times$10$^5$ DN. Most of these events would only be visible in cooler passbands including AIA 171 \AA\ and 193 \AA\ instead of hotter passbands. This means that it is not feasible to use the Differential Emission Measure \citep[DEM,][]{Hannah2012,Cheung2015}, which needs to utilise most if not all the optically thin AIA passbands, to estimate the plasma density and temperature, and thus the thermal energy of these jets. However, since the DEM is derived from the intensities of the optically thin passbands, with the intensity at one given temperature range proportional to the DEM at the same temperature range \citep{Cheung2015}, the above difficulties of the lack of signatures in hot EUV passbands can be bypassed by studying whether there are linear relationships between the AIA 304 \AA\ intensities and the AIA 171/193 \AA\ intensities of the coronal jets.

\begin{figure*}[htb]
\begin{center}
\includegraphics[width=0.7\hsize]{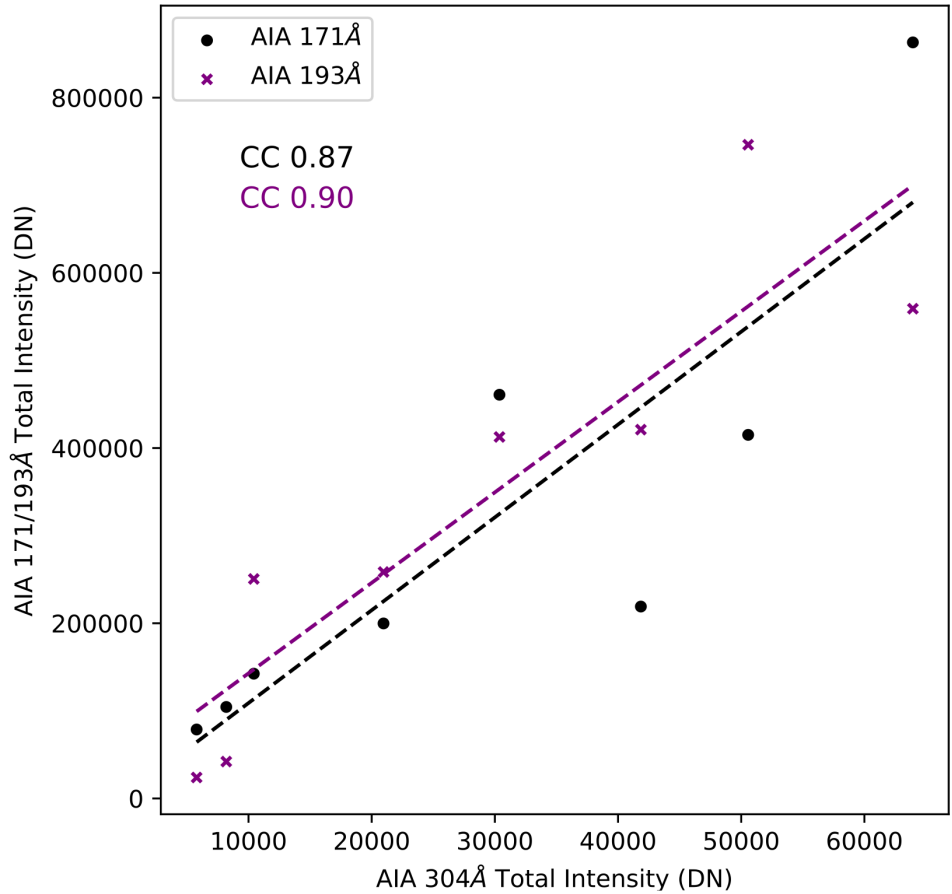}
\caption{\textbf{Intensities of the sample jets in different AIA passbands}. The $x$-axis is the AIA 304 \AA\ total intensity of eight sample jets from the dataset, and the $y$-axis represents their corresponding AIA 171 \AA\ (black dots) and 193 \AA\ (purple crosses) intensities. The black and purple dashed lines are the linear fits for the black dots and purple crosses, respectively.} \label{ext_fig_intensity}
\end{center}
\end{figure*}

Figure~\ref{ext_fig_intensity} shows the relationship between the AIA 304 \AA\ intensities and the AIA 171 \AA\ intensities (black dots), as well as the AIA 193 \AA\ intensities (purple crosses) of eight coronal jets from the 20 randomly chosen events mentioned in the previous section and listed in Table~\ref{tb_visi}. These events (rows 1-8 in Table~\ref{tb_visi}) have 304 \AA\ total intensities less than 6.5$\times$10$^5$ DN. Clear linear relationships could be found, with correlation coefficients (CCs) of 0.87 and 0.90 for the AIA 171 \AA\ and 193 \AA\ passbands, respectively. The fitted intensities (dashed lines in Fig.~\ref{ext_fig_intensity}) in the AIA 171 (193) \AA\ passband are on average 1.05$\pm$0.42 (1.58$\pm$1.19) times of the observed values. This indicates that the estimated thermal energies of these jets should be accurate in terms of the order of magnitude, with an average uncertainty of up to 60\% (grey shadow in Fig.~\ref{fig_power}). 

Nevertheless, the above power-law indices (-1.6 to -1.4) of jets' energy-frequency distribution are similar to that predicted by the volume model of solar flares \citep{Aschwanden2004}. Black curves in Figure~\ref{fig_power}b) are taken from Fig. 9 in \cite{Shibayama2013}, and show the power-law indices between $-1.8$ and $-1.5$ of nanoflares in the Sun to superflares in Sun-like stars with an energy range from $10^{24}$ erg to $10^{38}$ erg. The power-law distribution of the thermal energy of jets mostly falls within the range of that of microflares. The analogous power-law distributions of jets and flares are further evidenced by the alignment of the blue and black lines. Analogous power-law indices ($-2.1$ to $-1.5$) have also been previously found for the mass-/energy-frequency distribution of solar CMEs \citep[purple dashed line in Figure~\ref{fig_power}b,][]{Aschwanden2016,Lamy2019} and stellar CMEs \citep[e.g.,][]{Aarnio2012,Odert2017}.

The analogous power-law distributions between jets and flares are surprising given the vastly different plasma properties where jets and flares are triggered and the different magnetic topologies hosting their corresponding magnetic reconnections \citep{Shibata2007, Lin2000}. Moreover, about 56\% of the analysed jets are shorter than 30 Mm. Short jets are usually classified as macrospicules \citep{Kiss2017, Loboda2019}. They are not necessarily triggered by magnetic reconnections but could have been by alternative mechanisms including e.g. granular buffeting \citep{Roberts1979}, ion-neutral collision \citep{Haerendel1992} and $p$-mode driven shock waves \citep{Pontieu2004}.  Thirdly, many jets in our dataset have been found to experience some rotational motions. The thermal energies of rotating jets have been found related to not only magnetic reconnections (non-ideal) but also the gradual magnetic relaxation processes (ideal) afterwards \citep{Shibata2007, Raouafi2016}.

On the other hand, although the occurrence, intensity, and location of flares undergo significant change during a solar cycle,  it was found that there is not much difference between the power-law indices of flares during the solar maximum and minimum \citep{Bai2003}. The question is then: how are coronal jets modulated by the solar cycle and does their power-law index keep invariable at different stages of the solar cycle? To explore the answer to the above questions, the modulation of coronal jets by the solar cycle in terms of their spatial and temporal evolution was studied and is detailed in the following two subsections.

\subsection{Butterfly Diagram} \label{sec:butter}
First, let us revisit the spatial distribution of jets in Figure~\ref{fig_example}b). Some key patterns of the distribution of jets could be identified: 1) coronal jets that originate from lower latitudes seem to be slightly longer than those from the polar regions (with average lengths of 38 Mm and 34 Mm respectively); 2) there are more jets between $\pm$20\degree\ than at other non-polar regions; and 3) more jets are detected in the southern than the northern polar region.

\begin{figure}[htb]
\centering
\includegraphics[width=\hsize]{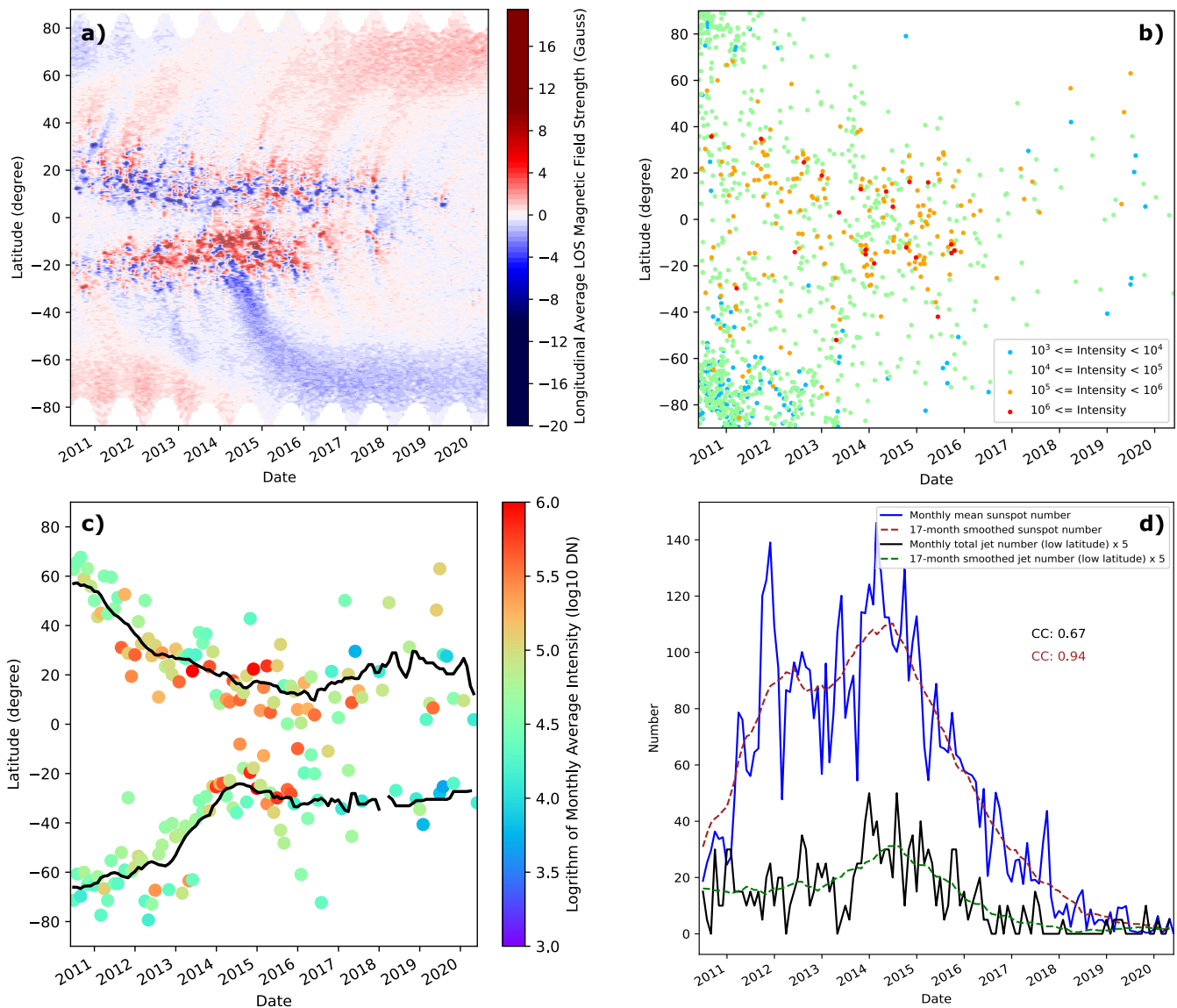}
\caption{\textbf{Butterfly diagrams of solar magnetic field and off-limb coronal jets.} \textbf{a)} is the longitudinal monthly average line-of-sight magnetic field strength observed by SDO/HMI, with red and blue colours denoting the positive and negative polarities, respectively. \textbf{b)} is the latitude and total intensity (in different colours) of all detected off-limb jets from 2010 June 1 to 2020 May 31. \textbf{c)} depicts the monthly average latitude and intensity (in different colours) of all detected off-limb jets. Black curves in this panel are the 17-month rolling averages of the monthly average latitudes for jets in the northern and southern hemispheres, respectively. The blue and brown curves in \textbf{d)} are the monthly mean sunspot number and its 17-month rolling average, respectively. The black and green curves are the monthly total jet number within the $\pm$20\degree\ latitudinal region and its 17-month rolling average, respectively. CC in black (brown) is the correlation coefficient between the black and blue (brown and green) curves.}
\label{fig_butterfly}
\end{figure}

The possible reasons for the above patterns are then explored via a more detailed study of the magnetic field distribution covering a length of a decade from June 2010 to June 2020. Figure~\ref{fig_butterfly}a) depicts the longitudinal monthly average line-of-sight magnetic field strength observed by the Helioseismic and Magnetic Imager (HMI) onboard SDO, with red and blue colours denoting the positive and negative polarities, respectively. Figure~\ref{fig_butterfly}b) shows the location and the number of all jets detected. Active regions (represented by strong magnetic field in Figure~\ref{fig_butterfly}a) exist with average latitudes mainly between $\pm$20\degree\ and are related in both time and space to more intense jets as revealed by red dots in Figure~\ref{fig_butterfly}b). This may explain why jets from lower latitudes are on average longer than those from polar regions. It may also explain why there have been more jets between $\pm$20\degree\ than other non-polar regions, considering that many jets are magnetic reconnection driven.

\begin{figure*}[htb]
\begin{center}
\includegraphics[width=0.9\hsize]{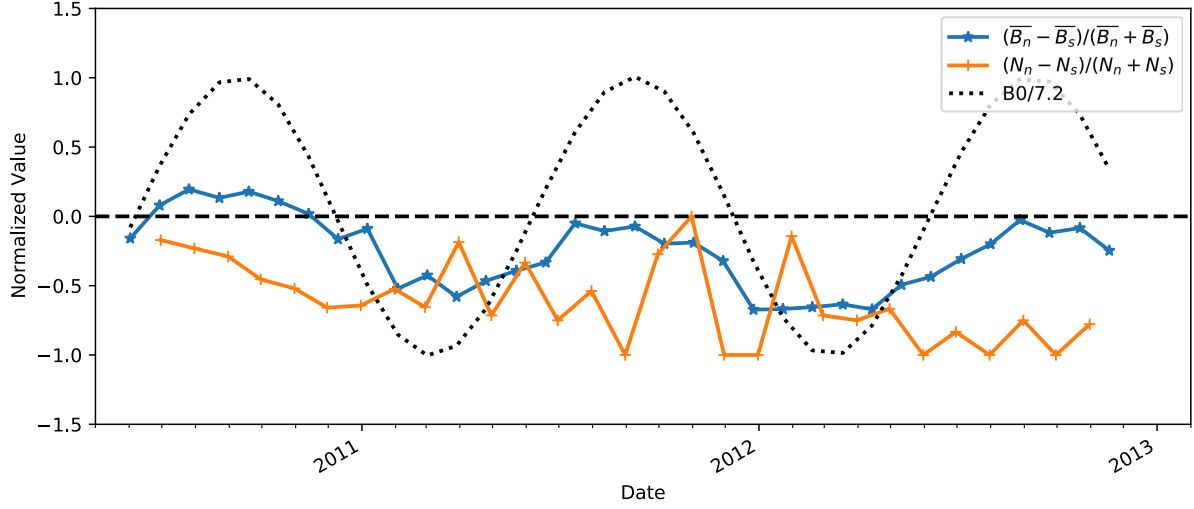}
\caption{\textbf{North-south asymmetry of the magnetic field strength and jet counts in polar regions}. $\overline{B_n}$ and $\overline{B_s}$ are the average absolute line-of-sight magnetic field strength between latitudes of 60\degree\ to 80\degree and -80\degree\ to -60\degree, respectively. The north-south asymmetry of the magnetic field $(\overline{B_n} - \overline{B_s})/(\overline{B_n} + \overline{B_s})$ is represented by the solid blue curve.  $N_n$ and $N_s$ are the monthly total jet numbers with latitudes above 60\degree\ and below -60\degree, respectively. The north-south asymmetry of the monthly total jet number $(N_n - N_s)/(N_n + N_s)$ is represented by the solid orange curve. The black dotted curve is for the normalised solar B0 angle of the solar disk centre seen from the Earth.} \label{ext_fig_asymm}
\end{center}
\end{figure*}

It is noticed from Figure~\ref{fig_example}b) that there have been apparently more jets in the southern than the northern pole, which could also be seen from the difference between the number of jets in the polar regions from 2010 to 2013 shown in Figure~\ref{fig_butterfly}b). This north-south asymmetry revealing more eruptive activities in the southern hemisphere is consistent with previous findings from sunspot areas and X-ray flares \citep[e.g.,][]{Janardhan2018,Prasad2020,Javaraiah2021} in Solar Cycle 24. The solid orange curve in Figure~\ref{ext_fig_asymm} shows the normalised difference of jet monthly counts detected in the northern (with latitudes above 60\degree) and the southern (with latitudes less than -60\degree) polar regions. The difference is always negative, again showing that there have been more jets in the southern than the northern polar region. 

The solid blue curve in Figure~\ref{ext_fig_asymm} depicts the north-south asymmetry of the polar magnetic field, with $\overline{B_n}$ and $\overline{B_s}$ as the monthly average absolute line-of-sight magnetic field strengths between latitudinal regions of 60\degree\ to 80\degree\ and -80\degree\ to -60\degree, respectively. The upper limits of the absolute latitudes are set to 80\degree\ because of the missing data between 80\degree\ (-80\degree) and the north (south) pole, which is a direct result of the annual variation of the B0 angle. The B0 angle represents the latitude of the disk centre in the Heliographic coordinate observed from the Earth and varies sinusoidally between $\pm$7.2\degree\ through the year as shown by the dotted curve in Figure~\ref{ext_fig_asymm}. $(\overline{B_n} - \overline{B_s})/(\overline{B_n} + \overline{B_s})$ oscillates annually along with the B0 angle with a CC of 0.82 between them. However, it is below zero in most cases even when the south pole was not visible from the Earth. This firmly shows that there have been stronger magnetic fields in the southern than in the northern polar region between 2010 and 2013.

A dominant feature of the distribution in Figure~\ref{fig_butterfly}b) is the migration of jet locations from higher to lower latitudes before 2015, especially the more intense jets, forming a butterfly diagram similar to that of the synoptic magnetic map in Figure~\ref{fig_butterfly}a). The jet butterfly diagram is more apparent in the monthly average latitudes and their 17-month average curves in Figure~\ref{fig_butterfly}c).  A direct comparison of Figure~\ref{fig_butterfly}a) and c) shows how well the migrations of the magnetic field and coronal jets are co-aligned in time. The simultaneous migrations, starting from around 2010 and ending around 2015, are consistent with the rising phase of Solar Cycle 24, as is indicated by e.g. the monthly mean sunspot number (blue curve in Figure~\ref{fig_butterfly}d) and its 17-month rolling average (brown curve in Figure~\ref{fig_butterfly}d). Black and green curves in Figure~\ref{fig_butterfly}d) are the monthly average number of jets detected within $\pm$20\degree\ latitude and its corresponding 17-month rolling average, both well correlated with their sunspot counterparts with correlation coefficients of 0.67 and 0.94, respectively.

\subsection{Quasi-annual oscillations} \label{sec:oscillaton}
\begin{figure}[htb]
\centering
\includegraphics[width=0.8\hsize]{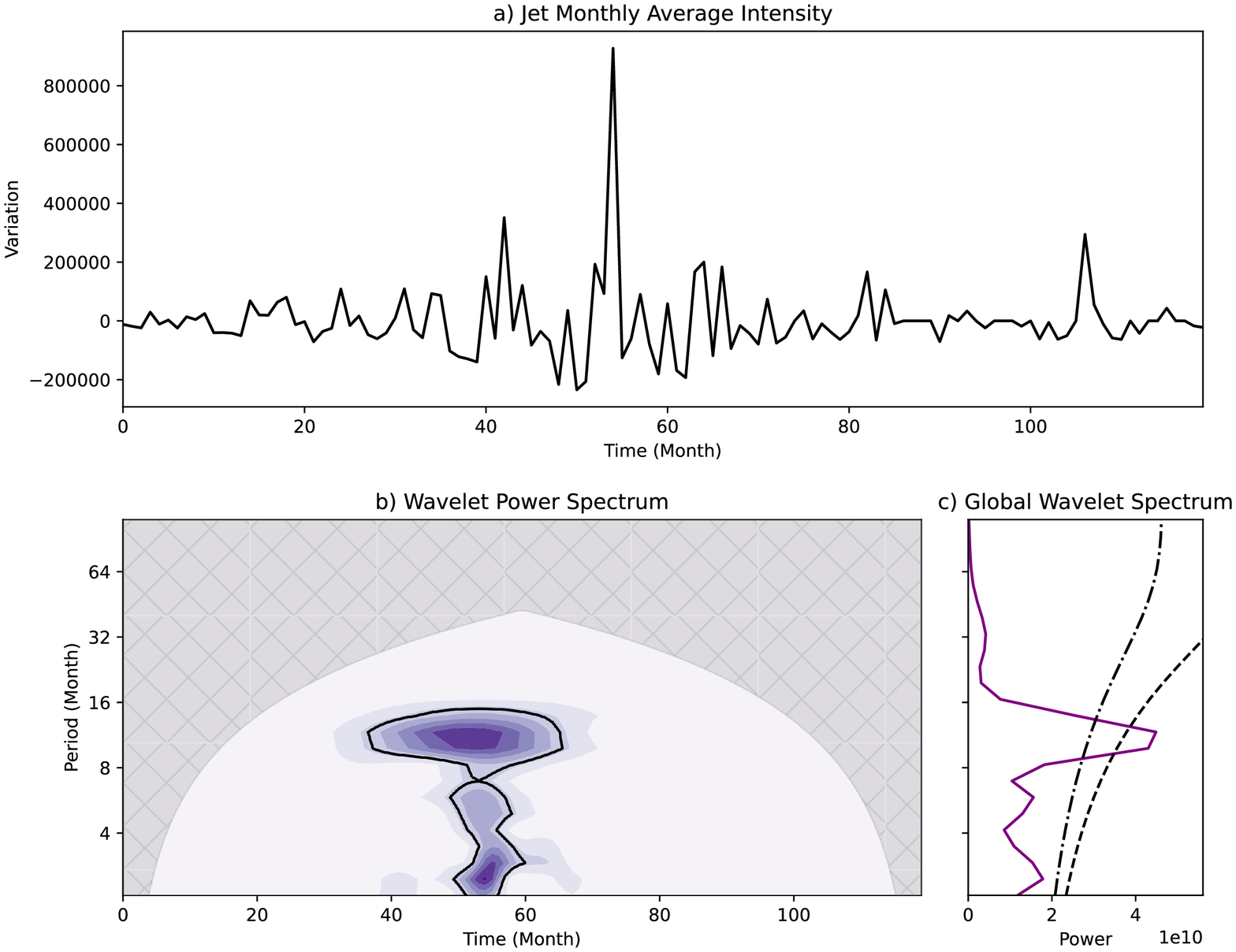}
\caption{\textbf{Wavelet spectra of the jet monthly average total intensity.} The black curve in \textbf{a)} is the local variation of the signal after removing the 17-month rolling average of the jet monthly average total intensity. The grey net in \textbf{b)} depicts the cone-of-influence \citep{Torrence1998} of the wavelet spectrum generated from the black curve in \textbf{a)}. Purple colours are local peaks of the wavelet spectrum with the contours as their corresponding 99\% confidence levels. The purple curve in \textbf{c)} is the global wavelet spectrum, with the dashed (dash-dotted) curve as its corresponding 99\% (95\%) confidence level.}
\label{fig_intensity}
\end{figure}

After removing the 17-month rolling average, the variation of the jet monthly average total intensity shows some periodic patterns (Figure~\ref{fig_intensity}a). 17 months were chosen because it is above the upper limit of the found periodicity ($\sim11$ months, Fig.~\ref{fig_intensity}). In addition, 17 months are not far away from the found periodicity which could avoid weakening the real periodicity and minimising the influence of the introduced artificial periods correlated to the window size. The periodicity can be then further confirmed by the wavelet \citep{Torrence1998} power spectrum with the ``Morlet'' mother function shown in Figure~\ref{fig_intensity}b), with purple colours denoting higher wavelet powers and black curves the local 99\% confidence levels. A dominant period between 8 to 16 months is seen outside the cone of influence (grey crosses) around the solar maximum, consistent with previous observations \citep{Bazilevskaya2000} of a number of solar indices in Solar Cycles 21 and 22. The global wavelet power (purple curve in Figure~\ref{fig_intensity}c) peaks at 11$\pm$2 months which is above both the 95\% and 99\% confidence levels (dash-dotted and dashed lines). A number of different window sizes of the rolling average from 9 to 21 months were tested and this peak period between 8 to 16 months exists in all cases.

\begin{figure*}[t!]
\begin{center}
\includegraphics[height=0.75\vsize]{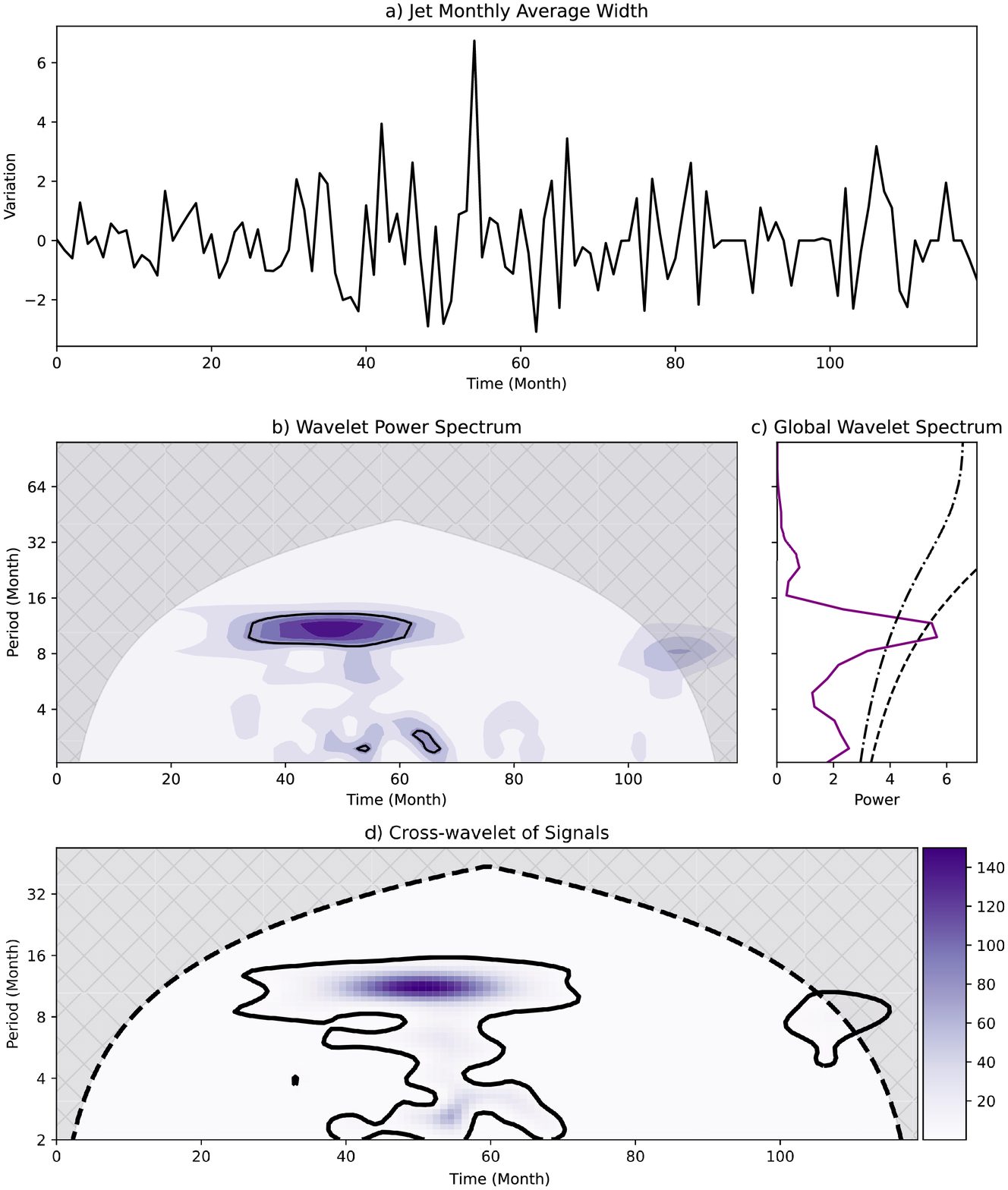}
\caption{\textbf{Wavelet spectra of the jet monthly average width.} The black curve in \textbf{a)} is the local variation of the signal after removing the 17-month rolling average of the jet monthly average width. The grey net in \textbf{b)} depicts the cone-of-influences of the wavelet spectrum generated from the black curve in \textbf{a)}. Purple colours are local peaks of the wavelet spectrum with the contours as their corresponding 99\% confidence levels. The purple curve in \textbf{c)} is the global wavelet spectrum, with the dashed (dash-dotted) curve as its corresponding 99\% (95\%) confidence level. \textbf{d)} shows the cross wavelet power spectrum between the jet monthly average width and jet monthly average total intensity. Solid black curves are the local 95\% confidence levels.}
\label{ext_fig_width}
\end{center}
\end{figure*}

\begin{figure*}[t!]
\begin{center}
\includegraphics[width=0.9\hsize]{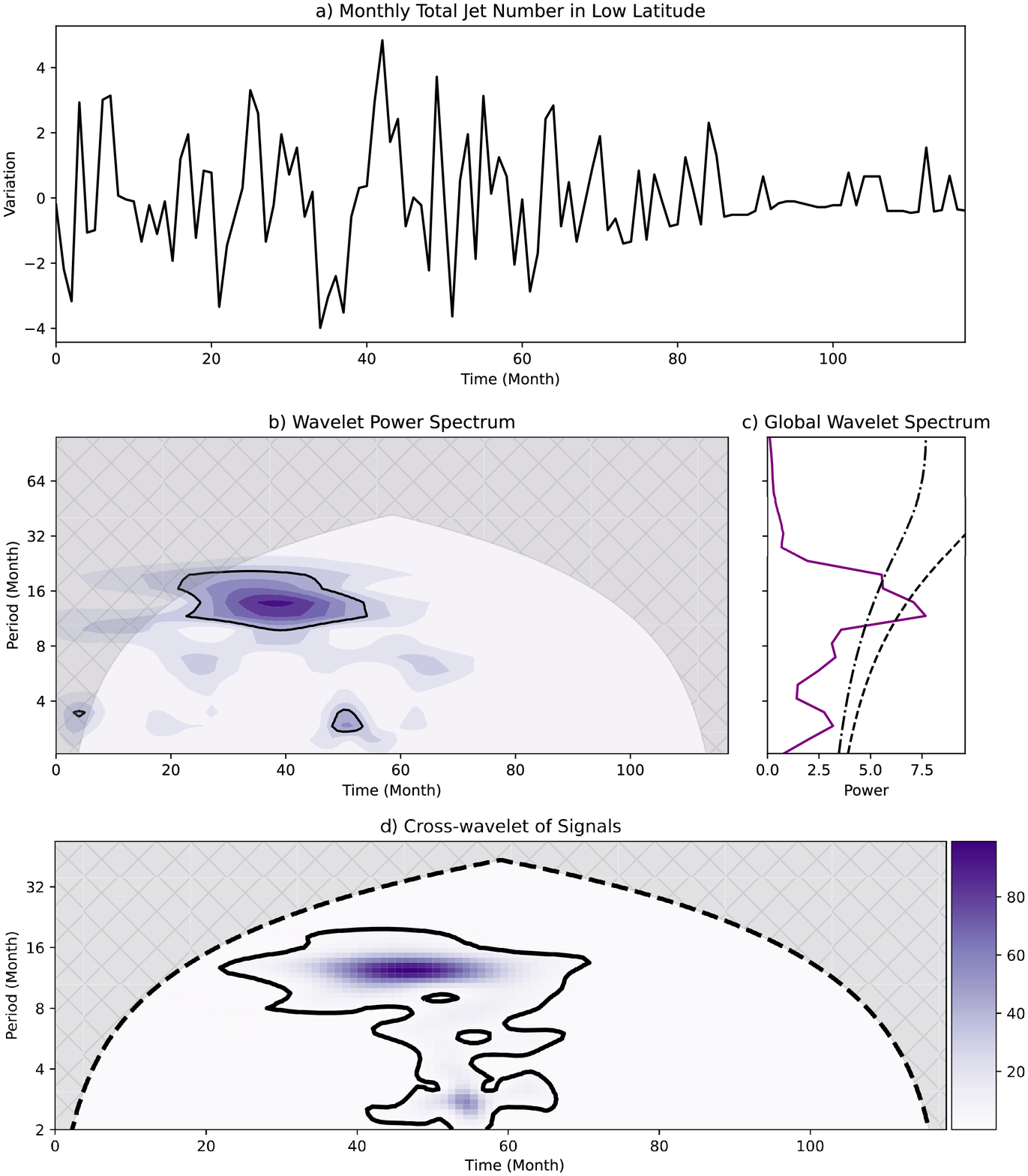}
\caption{\textbf{Wavelet spectra of the monthly total jet number in low latitudes between $\pm$20\degree.} Annotations are the same as those in Fig.~\ref{ext_fig_width}. } \label{ext_fig_count}
\end{center}
\end{figure*}

\begin{figure*}[t!]
\begin{center}
\includegraphics[width=0.9\hsize]{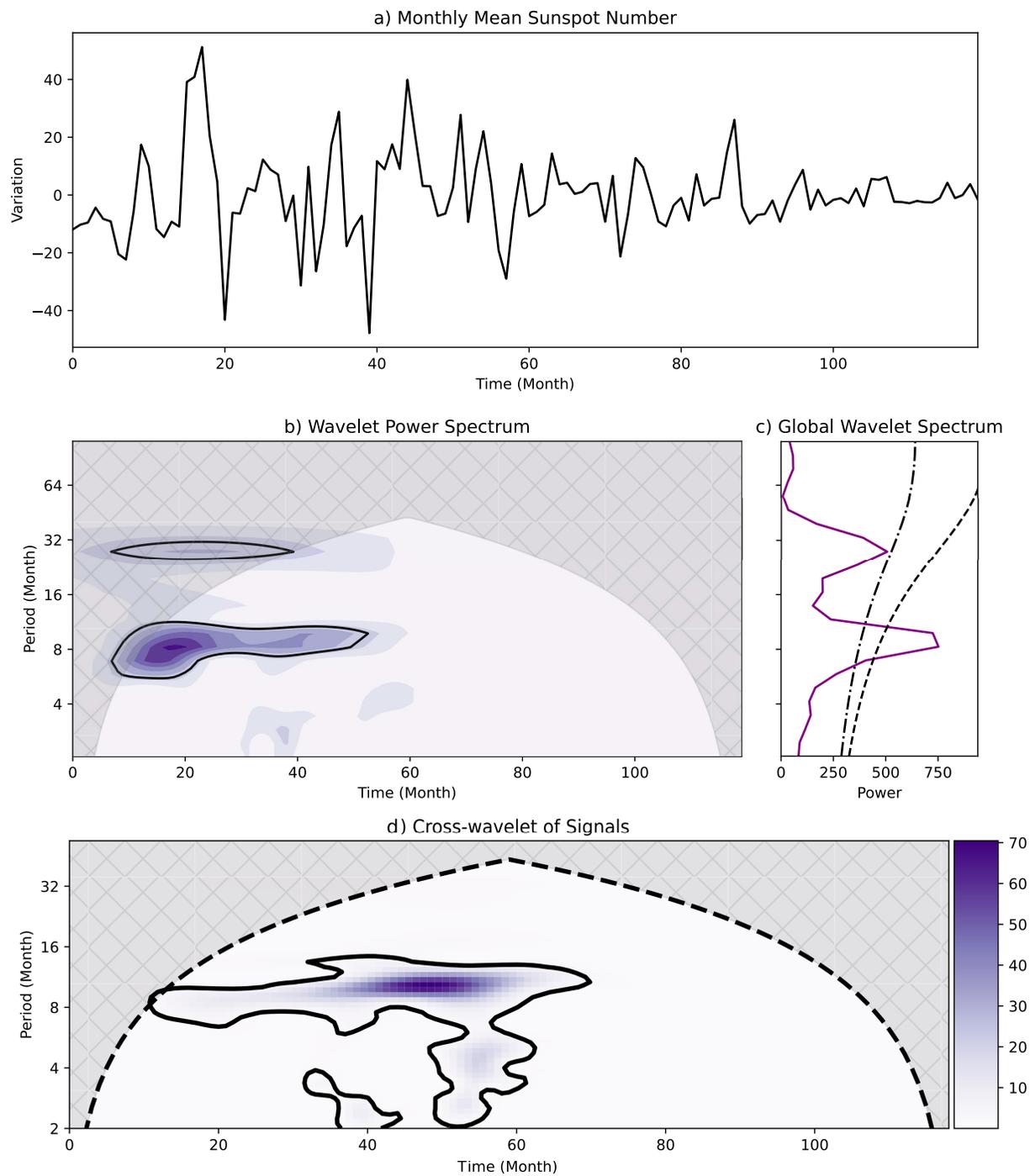}
\caption{\textbf{Wavelet spectra of the monthly mean sunspot number.} Annotations are the same as those in Fig.~\ref{ext_fig_width}. \label{ext_fig_sunspot}}
\end{center}
\end{figure*}

This quasi-annual periodicity is also present in all the other studied parameters of the detected jets, including their monthly average length, monthly average width (Fig.~\ref{ext_fig_width}), monthly average area, monthly average absolute latitude and monthly total count in low-latitude regions (between $\pm$20\degree, Fig.~\ref{ext_fig_count}). All peak global wavelet powers are above their corresponding 99\% confidence levels, except for the monthly average length that is above the 95\% but slightly below the 99\% confidence level. This quasi-annual oscillation also exists in the monthly mean sunspot number variation (Fig.~\ref{ext_fig_sunspot}). The high cross-wavelet powers shown in panel d) of Figures~\ref{ext_fig_width}-\ref{ext_fig_sunspot} further confirm the simultaneous quasi-annual oscillation of all the studied signals. 

It is worth noting that quasi-annual periods have been reported in the literature, through investigations of the corona and flares \citep{Deng2015,Gyenge2016,Singh2018}, chromospheric plages \citep{Xiang2019}, sunspots \citep{McIntosh2015,Mei2018}, coronal mass ejections \citep{Lou2003}, 10.7{\,}cm radio flux \citep{Lean1989}, solar wind parameters \citep{Chowdhury2012,Elborie2020}, geomagnetic activities \citep{Mursula2003} and even in a solar-like stellar system \citep{Massi2005}. The exact physical origin of these short-period variations remains unknown, while their period is consistent with that of the tachocline oscillations \citep{Howe2000}.  From this and employing MHD modelling, it was recently suggested that these observed quasi-annual oscillations might result from the interaction between solar magnetic fields, Rossby waves, and the tachocline \citep{Dikpati2017}. This, together with the less significant period of around 5 to 6 months in Fig.~\ref{fig_intensity}b),  reminds us of the well-known Rieger period \citep{Rieger1984,Bai1987,Bai1991} of $\sim$155 days which were originally observed from solar flares and later found in some other solar and interplanetary features \citep{Lean1989,Droge1990,Kudela2010,Chowdhury2012,Elborie2020}. The quasi-annual oscillations found in this work are most significant during the solar maximum, very similar to the behaviour of Rieger-type periods \citep{Rieger1984,Droge1990,Oliver1998,Lou2000}. Theories have also suggested Rieger-type periods as a result of the magnetic Rossby waves \citep{Lou2000,Zaqarashvili2010} with the Rieger periods (quasi-annual oscillation) corresponding to the number of nodes in the azimuthal direction of 12 to 13 (25 to 28). The above results together indicate that the quasi-annual oscillation of coronal jets could possibly be related to the 155-day Rieger period, which has been suggested as a result of the global solar dynamo \citep{Bai1991}.

The global wavelet power of the monthly mean sunspot number (Fig.~\ref{ext_fig_sunspot}) shows a relatively weak second peak (close to the 95\% confidence level) at around 2 years. This might indicate the existence of the quasi-biennial periodicity, which could possibly be related to the quasi-annual period. However, why this quasi-biennial periodicity is not present in the jet parameters studied (see e.g., Fig.~\ref{fig_intensity}, Fig.~\ref{ext_fig_width} and Fig.~\ref{ext_fig_count}) remains an open question.

It is also worth noting that the difference between jet counts of the northern and southern polar regions shows no simultaneous oscillation with the B0 angle (orange the black curves in Fig.~\ref{ext_fig_asymm}), with a CC of 0.004 between them. This suggests that the quasi-annual oscillations found above should not be resulting from the revolution of the Earth around the Sun. Moreover, if they were, the wavelet powers of the quasi-annual oscillations should have peaked between 2010 to 2013 when there were most samples, instead of between 2013 to 2015 around the solar maximum as one can see from Figure~\ref{fig_intensity} and Figures.~\ref{ext_fig_width}-\ref{ext_fig_count}.

Moreover, for further validation, SAJIA was applied to the same dataset again but with a lower limit used for the normalisation as the 95th percentile value of the ascendantly sorted array of each masked image. Of the 1215 confirmed jets found above, 892 were found to be included in the new collection of candidates. Repeating the performed analysis of these 892 jets has revealed very similar results supporting all the major findings detailed above, apart from that the peak global wavelet power of some signals (width and area) were found to be lower than their corresponding 95\% confidence levels.

From our analysis, one can find that the properties of jets are modulated by the solar cycle in a way similar to that of flares (in terms of the butterfly diagram and the quasi-annual oscillations). However, there is no significant difference in the power-law indices of coronal jets observed during the solar minimum (2010-2012 and 2015-2020) and the solar maximum (2012-2015). The MLE and least-square fit power-law indices are -1.46$\pm$0.12 and -1.72$\pm$0.14 during the solar minimum, and -1.53$\pm$0.17 and -1.71$\pm$0.21 during the solar maximum, respectively. This, again, agrees with the invariable property across solar cycles of the power-law index for solar flares. The above results together suggest that coronal jets, like flares, could also be avalanches of smaller events. Such a scenario is supported by high-resolution observations showing the existence of continuous sub-jets within a major eruption \citep{Liu2014} and subarcsecond blobs possibly as a result of plasmoid instability during the magnetic reconnection \citep{Zhang2019}.

\section{Discussons and Conclusions} \label{sec:conc}

\subsection{Coronal jets and photospheric magnetic activity}
In this paper, the newly developed Semi-Automated Jet Identification Algorithm (SAJIA) for detecting off-limb jets from full-disk solar observations was introduced. SAJIA has been applied to SDO/AIA 304 \AA\ observations obtained between June 2010 to June 2020, covering almost the whole range of Solar Cycle 24. 1215 off-limb coronal jets have been identified and confirmed, with their properties including the location, length, width, area, and intensity determined. It is found that spatial and temporal distributions of these jets are highly associated with photospheric magnetic activities, e.g.: 

\begin{enumerate}[1)]
\item there are more jets around the equator than at other non-polar regions; 
\item more jets are detected in the southern than the northern polar region, which is highly consistent with the north-south asymmetry of the polar magnetic field during Solar Cycle 24; 
\item the latitudes of jets migrate toward the equator from higher latitudes from the beginning to the end of the solar cycle thus forming a jet butterfly diagram; and 
\item properties of jets experience quasi-annual oscillations with an average period around 11 months, which is among the Rieger-type periods and has been found from many different types of solar activities including flares and CMEs.
\end{enumerate}

As far as we are aware, it is the first time that the latter two findings 3) and 4) have been shown.

We demonstrate that the modulation of coronal jets by the solar cycle is not only on the latitudinal behaviour and physical parameters of the jets studied above but also on the longitudinal distribution of these localised events, manifested by the ``active longitudes''  of jets. The term ``active longitudes" refers to longitudinal regions or belts that are more active than other regions on the Sun and cool stars. Active longitudes of sunspot groups, active regions, and flares have been widely studied since the discovery of the non-uniform longitudinal distribution of sunspots back to centuries ago, and have been suggested to be an important non-axisymmetric constraint for solar dynamo theories \citep{Losh1939,Bai1990hot,Bai2003,Usoskin2007,Gyenge2016}. Finding the existence of active longitudes of coronal jets could provide further evidence of the important roles the solar cycle plays in modulating not only the latitudinal but also longitudinal locations of jets. 

For a detected off-limb jet, its approximate Carrington longitude is determined by subtracting or adding 90\degree\ (depending on whether the jet was observed off the east or the west limb) to the Carrington longitude of the central meridian at the time of the observation. There will be uncertainties because an off-limb jet could have originated from a location slightly away ($\leq$ 15\degree, see Eq.~\ref{eq_cost}) from the limb. But these uncertainties are minor because the average width of active longitudes has been found \citep{Gyenge2016} to be as large as 60\degree. Starting from June 2010 (Carrington rotation CR2097), all jets observed in every two successive Carrington rotations are put into one group and their longitudinal distributions with 18 bins (corresponding to a bin size of 20\degree) are visualised following literature \citep{Gyenge2016}. To minimize statistical bias, all groups which have less than 18 events are omitted.

\begin{figure*}[htb]
\begin{center}
\includegraphics[width=0.9\hsize]{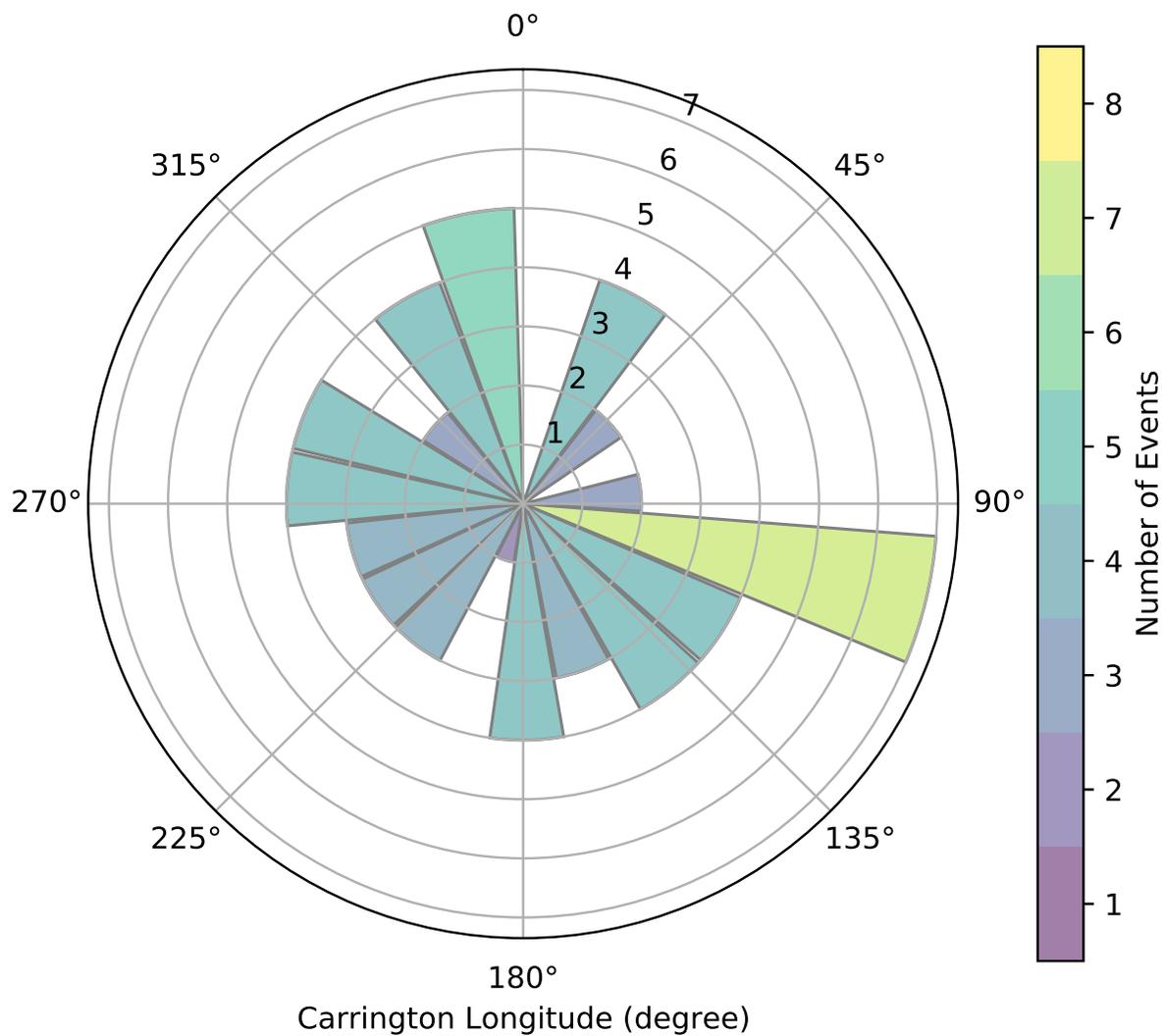}
\caption{\textbf{Longitudinal distribution of coronal jets in Carrington rotations CR2103 and CR2104}. Concentric circles and colours both represent numbers of events.\label{ext_fig_activelon}}
\end{center}
\end{figure*}

Therefore, histograms of the longitudinal distributions of jets detected in 17 Carrington rotations from June 2010 to July 2012 have been generated (Carrington rotations containing less than 18 events were omitted). For each of the distributions, the significance of their peaks is further investigated if there is any. With a strict definition that only considers a peak as significant if it is above 2$\sigma$ of the average value in each distribution, it was found that, of 17 Carrington rotations, 12 show evidence of the existence of active longitudes.  Figure~\ref{ext_fig_activelon} shows the longitudinal distribution of jets detected in CR2103 and CR2104. A peak number of events located at Carrington longitudes of 100\degree\ to 120\degree\ can be clearly seen. It is still an open question why active longitudes are not found in the other 5 Carrington rotations. To answer the above questions, a larger dataset will be constructed by applying SAJIA to SDO/AIA observations with an image cadence higher than 6 hours.

\subsection{Frequency distributions and coronal heating}
 Power-law distributions were found for both the solar coronal jet intensity and energy frequency distribution. The thermal power-law distribution remains almost invariable across the solar cycle despite the properties of jets being found to be significantly modulated by it. The thermal energy power-law index for jets is shown to be similar to thermal/non-thermal energy index values that have been found for solar and stellar flares, and CMEs, although they are triggered by different apparent processes with different magnetic topologies. These similar invariable power-law distributions of different types of eruptive events, covering a wide energy range across more than 14 orders of magnitude, i.e., nanoflares ($10^{23} - 10^{26}$~erg), coronal jets ($10^{26} - 10^{29}$~erg), solar flares and CMEs ($10^{27} - 10^{33}$~erg), and stellar flares and CMEs ($10^{33} - 10^{38}$~erg), strongly suggest that the same nonlinear statistics of scale-free processes are accountable for these eruptive events in both solar and stellar atmospheres.

It remains an open and fundamental question in solar and stellar physics of what the mechanism(s) may be that could act as the common underlying scale-free process behind the various apparently independent triggers, including the rather distinct ones of e.g., magnetic reconnections, MHD waves and localised bulk plasma motions, of solar and stellar eruptive events. Considering the results found from our research, we suggest that the solar dynamo is responsible ultimately for the universal power-law character. In other words, it is ultimately the solar dynamo that drives all these distinct formations of mass, momentum, and energy transport. The different eruptive phenomena, i.e., jets, flares, and CMEs, from different stars lying on the same power-law index might further suggest that these eruptions can be treated as the microscopic consequences of those solar dynamo driven global features from the bottom, especially when taking into account the observed power-law distributions of solar global features, including the global {\it p}-mode driven waves \citep{Tomczyk2009} and photospheric magnetic fluxes \citep{Parnell2009}. Substantial efforts will be needed to examine the above suggestion and to discover the details of the physics, with expertise in observations, numerical simulations, and analytic modelling. 

Most spicules \citep[with length less than 14 Mm,][]{Pontieu2004,Shibata2007,Samanta2019} were excluded from our semi-automated identification algorithm. It would be of great interest to extend our research down to the scale of spicules \citep[with typical potential energies $\leq 10^{25}$ erg,][]{Sterling2000}. Simplified estimations have suggested the power-law relationship between coronal jets and spicules as promising \citep{Sterling2016}. \cite{Hudson1991} demonstrated that small-scale events would dominate the heating if the corresponding power-law index of their energy-frequency distribution is less than -2. Observations of quiet-sun nanoflares revealed a power-law index less than -2 \citep{Parnell2000}, suggesting that events with the lowest energies may dominate the heating of corona in quiet-sun regions. However, \cite{Aschwanden2000} suggested that this higher power-law index (-2 v.s. -1.8 found by many other studies) could be a result of methodical differences. Nevertheless, both studies \citep{Parnell2000,Aschwanden2000} found that energies from nanoflares were not enough to balance the total energy losses in the corona. Therefore, we suggest that finding the power-law distribution and its index from the energy-frequency distribution of spicules would be significant in exploring the similarities and differences of their triggering mechanism with coronal jets and addressing the potential role of spicules in coronal heating.

\vspace{5mm}
\noindent \textbf{ACKNOWLEDGEMENT}
\vspace{3mm}

We thank the anonymous reviewer for their careful reading of our manuscript and insightful comments and suggestions. We acknowledge the use of the data from the Solar Dynamics Observatory (SDO). SDO is the first mission of NASA's Living With a Star (LWS) program. The SDO/AIA and SDO/HMI data are publicly available from NASA's SDO website (\url{https://sdo.gsfc.nasa.gov/data/}). Sunspot numbers are available at the Solar Influences Data analysis Center (SIDC) (\url{http://sidc.be/}), Royal Observatory of Belgium, Brussels. J.L. acknowledges support from the Frontier Scientific Research Program of Deep Space Exploration Laboratory (2022-QYKYJH-ZYTS-016), the NSFC Distinguished Overseas Young Talents Program, the Informatization Plan of Chinese Academy of Sciences (CAS-WX2022SF-0103), the Leverhulme Trust via grant RPG-2019-371, and the STFC under grant No. ST/P000304/1.  A.S. and Y.W. are supported by the Strategic Priority Program of the Chinese Academy of Sciences (Grant No. XDB41000000) and NSFC (No. 41774178). R.E. is grateful to STFC (UK, grant number ST/M000826/1), NKFIH OTKA (Hungary, grant No. K142987) and the Royal Society. R.E. also acknowledges the support received by the CAS Presidents International Fellowship Initiative Grant No. 2019VMA052 and the warm hospitality received at USTC, where part of his contribution was made. J.Z. was funded by NASA grants NNH17ZDA001N-HSWO2R, NNH17ZDA001N-LWS, NNH19ZDA001N-TMS. S.S. acknowledges the grant from project No. C1791784 implemented with the support provided by the Ministry of Innovation and Technology of Hungary from the National Research, Development and Innovation Fund, financed under the KDP-2021 funding scheme. 

%

\vspace{5mm}
\facilities{SDO(AIA and HMI)}

\clearpage
\bibliographystyle{aasjournal}
\bibliography{main.bbl}

\begin{table}[!ht]
\caption{Visibility of 20 randomly chosen coronal jets in AIA EUV passbands.}
\label{tb_visi}
    \begin{tabular}{l|l|l|l|l}
    \hline
        Event ID & Time (UT) & PA (degree) & Intensity (DN) & Visibility in AIA passbands  \\ \hline
        1417 & 2012/2/11 12:00 & 192.6  & 5788  & Hardly visible in 171 \AA\ or 193 \AA\ \\ \hline
        714 & 2011/2/22 18:00 & 173.8  & 8196  & Dark in 171 \AA\ and 193 \AA\  \\ \hline
        3693 & 2017/7/7 0:00 & 292.4  & 10430  & Bright in 171 \AA\ and dark in 193 \AA\  \\ \hline
        2919 & 2014/5/16 6:00 & 206.7  & 20961  & Bright in 171 \AA\ and dark in 193 \AA\  \\ \hline
        3652 & 2017/2/11 12:00 & 326.9  & 30373  & Bright in 171 \AA\ and dark in 193 \AA\  \\ \hline
        3326 & 2015/4/8 0:00 & 285.0  & 41853  & Bright in 171 \AA\ and dark in 193 \AA\  \\ \hline
        843 & 2011/4/15 6:00 & 41.5  & 50563  & Bright in 171 \AA\ and dark in 193 \AA\  \\ \hline
        3907 & 2019/10/7 0:00 & 278.9  & 63936  & Bright in 171 \AA\ and dark in 193 \AA\  \\ \hline
        3293 & 2015/3/13 18:00 & 118.6  & 81936  & Bright in 171 \AA\ and dark in 193 \AA\  \\ \hline
        1174 & 2011/9/30 0:00 & 63.4  & 104322  & Bright in both 171 \AA\ and 193 \AA\  \\ \hline
        3504 & 2015/12/8 6:00 & 276.0  & 202217  & Dark in both 171 \AA\ and 193 \AA\  \\ \hline
        1327 & 2011/12/6 12:00 & 285.4  & 308552  & Bright in both 171 \AA\ and 193 \AA\  \\ \hline
        3661 & 2017/4/4 12:00 & 285.1  & 422448  & Bright in both 171 \AA\ and 193 \AA\  \\ \hline
        2870 & 2014/4/19 12:00 & 276.1  & 501198  & Bright in all EUV passbands  \\ \hline
        1387 & 2012/1/16 18:00 & 273.0  & 602989  & \multicolumn{1}{p{6cm}}{Bright in 171 \AA\ and 131 {\AA}; dark in 193 {\AA}, 211 {\AA}, 335 \AA\ and 94 \AA\ } \\ \hline
        2013 & 2012/12/19 18:00 & 288.9  & 741102  & Bright feature in all EUV passbands  \\ \hline
        1324 & 2011/12/4 18:00 & 308.3  & 829421  & Bright in all EUV passbands  \\ \hline
        1161 & 2011/9/25 12:00 & 297.8  & 1040299  & \multicolumn{1}{p{6cm}}{Bright in 171 {\AA}, 131 {\AA}; dark in 193 {\AA}, 211 \AA\ and 335 {\AA}; not visible in 94 \AA\ }  \\ \hline
        3385 & 2015/6/12 12:00 & 227.4  & 2146707  & Bright in all EUV passbands  \\ \hline
        2637 & 2013/12/5 6:00 & 105.4  & 3911468  & Bright in all EUV passbands  \\ \hline\multicolumn{5}{p{0.92\textwidth}}{\footnotesize Note -- PA represents position angle in units of degrees. The characteristic temperatures for the 7 AIA EUV passbands (304 {\AA}, 171{\AA}, 193 {\AA}, 211 {\AA}, 335 {\AA}, 94 {\AA}, and 131 {\AA}) are 0.05 MK, 0.6 MK, 1.6 MK, 2.0 MK, 2.5 MK, 6.3 MK and 10 MK (0.4 MK), respectively \citep{Lemen2012}.}
    \end{tabular}
\end{table}

\clearpage

\end{document}